\documentclass[10pt]{article}
\usepackage{fullpage}
\usepackage{amsmath}
\usepackage{amsfonts}
\usepackage{amssymb}
\usepackage[dvips]{epsfig}
\usepackage{color}

\def\blue{\textcolor{black}}

\def\##1{\underline{#1}}
\def\=#1{\underline{\underline{#1}}}

\def\+#1{\underline{\bf #1}}
\def\*#1{\underline{\underline{\bf #1}}}

\def\~#1{{\tilde #1}}

\def\.{\mbox{ \tiny{$^\bullet$} }}

\def\eps{\epsilon}
\def\epso{\epsilon_{\scriptscriptstyle 0}}
\def\muo{\mu_{\scriptscriptstyle 0}}

\def\ko{k_{\scriptscriptstyle 0}}

\def\c#1{\cite{#1}}
\def\l#1{\label{#1}}
\def\r#1{(\ref{#1})}

\def\le{\left(}
\def\ri{\right)}
\def\les{\left[}
\def\ris{\right]}
\def\lec{\left\{}
\def\ric{\right\}}

\renewcommand{\thefootnote}{\fnsymbol{footnote}}

\begin{document}

\begin{center}

\LARGE{ {\bf Towards optical sensing with hyperbolic metamaterials}}
\end{center}
\begin{center}
\vspace{5mm} \large

 Tom G. Mackay\footnote{E--mail: T.Mackay@ed.ac.uk}\\
{\em School of Mathematics and
   Maxwell Institute for Mathematical Sciences\\
University of Edinburgh, Edinburgh EH9 3FD, UK}\\
and\\
 {\em NanoMM~---~Nanoengineered Metamaterials Group\\ Department of Engineering Science and Mechanics\\
Pennsylvania State University, University Park, PA 16802--6812,
USA}
\normalsize

\end{center}

\bigskip
\renewcommand{\thefootnote}{\arabic{footnote}}
\setcounter{footnote}{0}

\bigskip

\vspace{5mm}

\begin{center}
\vspace{1mm} {\bf Abstract}

\end{center}

\vspace{4mm}
A possible means of optical sensing,
based on a porous hyperbolic material
which is infiltrated by a fluid containing an analyte to be sensed,
 was investigated theoretically. The sensing mechanism relies on the observation
that extraordinary plane waves propagate in the infiltrated hyperbolic material only in directions enclosed by a cone aligned with the optic axis of the infiltrated hyperbolic material. The angle this cone subtends to the plane perpendicular to the optic axis is  $\theta_c$.
The sensitivity of $\theta_c$ to changes in refractive index of the infiltrating fluid, namely $n_b$, was explored; also considered were the permittivity parameters and porosity of the hyperbolic material, as well as the shape and size of its pores.
Sensitivity was gauged by the derivative
$d \theta_c / d n_b$. In parametric  numerical studies, values of $d \theta_c / d n_b$ in excess of 500 degrees per refractive index unit were computed, depending upon the constitutive parameters of the porous hyperbolic material and infiltrating fluid, and the nature of the porosity.
In particular, it was observed that exceeding large values of $d \theta_c / d n_b$ could be attained as the negative--valued eigenvalue of the infiltrated hyperbolic material approached zero.

\vspace{5mm}


\vspace{5mm}


\section{Introduction}

Through judicious design, engineered composite materials~---~sometimes referred to as metamaterials~---~can offer unique
opportunities for controlling the propagation of light.
Hyperbolic metamaterials represent a particularly interesting category of such
engineered composite materials \c{Kivshar}. The simplest example of a hyperbolic material, at least from a mathematical perspective, is provided
by a nondissipative uniaxial dielectric material whose permittivity dyadic is indefinite (i.e., it possesses both positive-- and negative--valued eigenvalues).
The
hyperbolic--material
concept may be extended to dissipative uniaxial dielectric materials whose permittivity dyadics possess real parts which are indefinite.
Such materials are promising for a host of potential applications, including as platforms for negative refraction (without concomitant negative--phase--velocity propagation) \c{Smith04,Schilling}, subwavelength imaging \c{Liu_Science_2007,Li_JAP,Lemoult},
 radiative thermal energy transfer \c{Hyperbolic_Planck,Hyperbolic_Guo_OE}, as analogs of curved spacetime \c{Smoly,ML_PRB}, and
as diffraction gratings which can direct light into a large number of refraction channels \c{Depine_NJP}, for examples.
Hyperbolic metamaterials may be fabricated as metal/dielectric composites, with the components arranged as layered sheets \c{Shen_PRB,Orlov} or as arrays of nanowires \c{Wire}. A simpler method of fabrication may be achieved via the homogenization of a random distribution of
electrically--small ellipsoidal particles \c{Depine_MOTL}. In this context it is notable that hyperbolic materials are also found in nature in certain spectral regimes; for example, graphite in the ultraviolet regime \c{Graphite}. Greater scope for hyperbolic metamaterials may be offered by dielectric--magnetic materials whose permittivity and permeability dyadics are both indefinite (or whose real parts are both indefinite) \c{Smith03}. A major challenge in developing hyperbolic metamaterials is to overcome dissipative losses. The use of  metals such as silver at visible wavelengths and/or active component materials may help to minimize such losses \c{Noginov,Vincenti}.

Our interest here lies in a novel optical--sensing application of hyperbolic metamaterials. We consider a porous hyperbolic material which is infiltrated by a fluid containing an analyte to be sensed. The presence of the analyte--containing fluid changes the optical properties of the hyperbolic material in manner that may be harnessed for optical sensing. Specifically, extraordinary plane waves propagate in the infiltrated hyperbolic material only in directions enclosed by a cone aligned with the optic axis of the
infiltrated
hyperbolic material. The angle this cone subtends to the plane perpendicular to the optic axis, namely $\theta_c$, may be acutely sensitive to the refractive index of the infiltrating fluid. Thus, by monitoring the magnitude of $\theta_c$, the concentration of analyte in the fluid may be gauged.
 In the following we explore the sensitivity of $\theta_c$ to changes in refractive index of the infiltrating fluid; also considered are the porosity and permittivity parameters of the hyperbolic material, as well as the shape and size of its pores. The theoretical approach adopted is based on the extended version \c{M_JNP} of the well--established Maxwell Garnett homogenization formalism \c{MG_MOTL,ML_PiO}, which requires that the
hyperbolic material is highly porous and that
the
pores are much smaller than the wavelengths under consideration.

As regards notation, in the following vectors are underlined, with the $\hat{}$ symbol denoting a unit vector. Thus, the unit vectors aligned with the Cartesian axes are $\lec \hat{\#x}, \hat{\#y}, \hat{\#z} \ric$. Double underlining indicates a 3$\times$3 dyadic (i.e., a  second--rank Cartesian tensor), with $\=I = \hat{\#x} \, \hat{\#x} + \hat{\#y} \, \hat{\#y} + \hat{\#z} \, \hat{\#z}$ being the identity dyadic. The permittivity and permeability of free space are written as $\epso$ and $\muo$, respectively; $\ko = \omega \sqrt{\epso \muo}$ is the free--space wave number, with $\omega$ being the angular frequency.
The real and imaginary parts of complex--valued quantities are delivered by the operators $\mbox{Re} \lec \. \ric$ and $\mbox{Im} \lec \. \ric $, respectively; and $i = \sqrt{-1}$.

\section{Plane waves in hyperbolic materials} \l{Hyperbolic_sec}

Let us consider a homogeneous uniaxial  dielectric material characterized by the
Tellegen constitutive relations \c{ML_PiO}
\begin{equation}
\left.\begin{array}{l}
\#D(\#r) = \=\eps \. \#E (\#r) \vspace{6pt} \\
\#B(\#r) = \muo \, \#H (\#r)
\end{array}\right\},
\l{crs}
\end{equation}
where the
 permittivity dyadic
\begin{equation} \l{eps_uniaxial}
\=\eps = \eps^\perp \le \=I - \hat{\#z} \,\hat{\#z} \ri + \eps^\parallel \hat{\#z} \, \hat{\#z}.
\end{equation}
 The optic axis of the material is parallel to
 the unit vector $\hat{\#z}$ \c{BW}. Largely we focus on nondissipative materials for which the permittivity parameters $\eps^\perp, \eps^\parallel \in \mathbb{R}$,
 but this restriction is relaxed in \S\ref{exMG_sec} wherein weakly dissipative materials (for which $\eps^\perp, \eps^\parallel \in \mathbb{C}$) are considered.

Now suppose that plane waves with field phasors
\begin{equation}
\left.\begin{array}{l}
\#E(\#r) = \#E_0\, \exp \le i \#k \. \#r  \ri \vspace{6pt} \\
\#H(\#r) = \#H_0\, \exp \le i \#k \. \#r \ri
\end{array}\right\}
\l{pw}
\end{equation}
propagate in the hyperbolic material characterized by Eqs.~\r{crs} and \r{eps_uniaxial}. In the case of uniform plane waves, propagating in a nondissipative material, the wave vector components are given as
\begin{equation}
\#k = k_x \, \hat{\#x} + k_y \, \hat{\#y} + k_z \, \hat{\#z} , \qquad (k_x, k_y, k_z \in \mathbb{R}).
\end{equation}
The  source--free Maxwell curl postulates
\begin{equation}
\left.\begin{array}{l}
\nabla \times \#E(\#r) - i \omega \#B ( \#r ) = \#0 \vspace{6pt}\\
\nabla \times \#H(\#r)  + i \omega \#D ( \#r ) = \#0
\end{array}\right\}
\l{Maxwell}
\end{equation}
combined with the constitutive relations \r{crs}
provides   the vector Helmholtz equation
\begin{equation}
\les \,\le \nabla \times \=I \,\ri \. \le \nabla \times \=I\, \ri
- \muo \omega^2 \=\eps\,\ris \.  \#E (\#r) = \#0\,, \l{Helmholtz}
\end{equation}
which yields the
 dispersion relation
\begin{equation}
\le k_x^2 + k_y^2 + k_z^2 - \eps^\perp \muo \omega^2 \ri \les \,
\le k_x^2 + k_y^2 \ri \eps^\perp +
k_z^2
\eps^\parallel  - \eps^\perp \eps^\parallel \muo \omega^2 \ris = 0\,
\l{disp}
\end{equation}
for plane waves specified by Eqs.~\r{pw}${}_1$. Two distinct plane--wave solutions emerge from Eq.~\r{disp}:  the \emph{ordinary}
plane wave whose wave--vector components satisfy \c{Chen}
\begin{equation} \l{or_dr}
k_x^2 + k_y^2 + k_z^2 =  \frac{\ko^2 \eps^\perp }{\epso} \,;
\end{equation}
and  the \emph{extraordinary}
plane wave whose wave--vector components satisfy \c{Chen}
\begin{equation} \l{ex_dr}
\frac{ k_x^2 + k_y^2 }{ \eps^\parallel} +
\frac{k_z^2}
{\eps^\perp}  =  \frac{\ko^2}{\epso} \,.
\end{equation}
Note that the ordinary and extraordinary plane waves can be readily distinguished from each other by their polarization states. For example, if propagation is restricted to the $xz$ plane, then the
 vector Helmholtz equation
\r{Helmholtz} yields
 the eigenvector solutions
 \begin{equation}
  \#E_0 = \left\{
  \begin{array}{lcr}
   E_y \,
\hat{\#y} && \mbox{ for the ordinary plane wave} \vspace{6pt}\\
  \displaystyle{E_{xz} \, \les \hat{\#x}   -  \frac{  \eps^\parallel \, k_z }{ \eps^\perp \,
k_x  } \, \hat{\#z} \,\ris } && \mbox{ for the extraordinary plane wave}
\end{array}
\right.
,
\end{equation}
 where the amplitudes $E_y$ and $E_{xz}$ may be determined from the boundary conditions.

Conventionally, in the realm of crystal optics  $\=\eps$ is positive definite (i.e., $\eps^\perp > 0$ and $\eps^\parallel > 0$);
therein, the ordinary dispersion relation \r{or_dr}   has a spherical representation in $\le k_x, k_y, k_z \ri$ space and
 the extraordinary dispersion relation \r{ex_dr} has a spheroidal representation in $\le k_x, k_y, k_z \ri$ space \c{BW}.
However,
our interest here is in the case where $\=\eps$ is indefinite.  To be specific, suppose that $\eps^\perp > 0$ and $\eps^\parallel < 0$.
 In this case,
the dispersion relations for the ordinary and extraordinary plane waves are dramatically different.
 The ordinary dispersion relation \r{or_dr}   has a spherical representation in $\le k_x, k_y, k_z \ri$ space
 but the extraordinary dispersion relation \r{ex_dr} has a hyperbolic representation in $\le k_x, k_y, k_z \ri$
 space. Accordingly,
 materials characterized by such an indefinite permittivity dyadic are known as hyperbolic materials \c{Kivshar}.

 An obvious distinction between the ordinary and extraordinary plane waves for hyperbolic materials is that ordinary plane waves can propagate in all directions whereas  extraordinary plane waves can only propagate in a restricted range of directions. Specifically, extraordinary plane wave directions are bounded by the cone
 \begin{equation} \l{cone}
 k_z^2 = \le k_x^2 + k_y^2 \ri \tan^2  \theta_c  ,
 \end{equation}
 which subtends the angle
 \begin{equation} \l{theta_c_def}
 \theta_c = \tan^{-1} \sqrt{ - \frac{\eps^\perp }{\eps^\parallel} }
 \end{equation}
 to the $k_x$-$k_y$ plane. The limiting values of the extraordinary  wave--vector  components are asymptotic to this cone.
 By way of illustration,
the extraordinary  wave--vector components $k_x $ and $k_z $ in the $k_y = 0$ plane  are plotted in Fig.~\ref{fig1} for the case where the permittivity parameters $\eps^\perp = 3.5 \epso$ and $\eps^\parallel = -2 \epso$.
 For wave--vector directions lying outside those prescribed by the cone \r{cone}, the extraordinary dispersion relation \r{ex_dr} yields only evanescent (i.e., nonpropagating) solutions. That is, extraordinary
plane waves with wave--vector component $k_z = k \cos \psi$, where $k = \sqrt{ \#k \cdot \#k}$, only propagate in the directions
given by $0 \leq \psi < \le \pi/2 \ri - \theta_c$. The transition  between propagating solutions and evanescent solutions that occurs at $\psi = \le \pi/2 \ri - \theta_c$ has the potential to be harnessed for optical sensing.

\section{Homogenization} \l{MG_Sec}

As a platform for optical sensing, let us consider a mixture of two  materials: material `a' is a porous hyperbolic  material as characterized in Eqs.~\r{crs} and \r{eps_uniaxial}; and material `b' is a fluid of refractive index $n_b$,  containing a  to--be--sensed  analyte. The permittivity dyadics  of the two component materials are written as
\begin{equation}
\left.
\begin{array}{l}
\=\eps_{\,a} = \eps^\perp_a \le \=I - \hat{\#z} \,\hat{\#z} \ri + \eps^\parallel_a \, \hat{\#z} \, \hat{\#z} \vspace{6pt} \\
\=\eps_{\,b} = \eps_b \, \=I
\end{array}
\right\},
\end{equation}
with $\eps_b = n_b^2 \epso$.
All material `a'  pores are assumed to have the same shape and orientation. Each pore
is taken to be spheroidal in shape,
with the rotational symmetry axis of each  pore being aligned with the optic axis of material `a'.
Thus, the surface of each component pore relative to its centroid
is traced out by the
 vector
\begin{equation} \l{rs_def}
\#r_{\,s}  = \rho \, \=U \. \hat{\#r},
\end{equation}
wherein $\rho > 0$ is a measure of the linear dimensions of the  pore; the shape--and--orientation dyadic
\begin{equation} \l{Udef}
\=U = U^\perp \le \=I - \hat{\#z} \,\hat{\#z} \ri + U^\parallel \, \hat{\#z} \, \hat{\#z}, \qquad \qquad \le U^\perp, U^\parallel > 0 \ri,
\end{equation}
and
$\hat{\#r}$ is the (unit) vector that prescribes the surface of the unit sphere
The pores are randomly distributed in material `a' with volume fraction $f_a$. Hence, the volume fraction of material `b' is $1-f_a$.
\blue{A schematic diagram of the two component materials is provided in Fig.~\ref{figx}.
The particular  nanostructure of the hyperbolic host material (at length scales smaller than the pores it contains)  could be engineered as a  one-dimensional \c{Wire}, two-dimensional \c{Shen_PRB,Orlov} or three-dimensional  \c{Depine_MOTL} structure; or indeed, in principle, this material could even be a
 natural material \c{Graphite}. However, the particular nanostructure of the hyperbolic host material (at length scales smaller than the pores it contains)
 is not directly pertinent  to the proceeding analysis.
}

Provided that the size parameter $\rho$ is much smaller than the wavelengths under consideration, the mixture of materials `a' and `b' may be regarded as being effectively homogeneous. The permittivity dyadic of the corresponding homogenized composite material (HCM), namely \c{ML_PiO}
\begin{equation}
\=\eps_{\,hcm} = \eps^\perp_{hcm} \le \=I - \hat{\#z} \,\hat{\#z} \ri + \eps^\parallel_{hcm} \, \hat{\#z} \, \hat{\#z} ,
\end{equation}
 can be estimated using the well--established Maxwell Garnett formalism. For the case where material `a' is highly porous, i.e., $f_a \lessapprox 0.3$, the Maxwell Garnett estimate is given by \c{ML_PiO}
 \begin{equation} \l{eMG}
\=\eps_{\,hcm} = \=\eps_{\,b} + f_a \, \=\alpha  \. \le \=I - f_a \=D_{\,s} \. \=\alpha \ri^{-1},
 \end{equation}
 wherein the polarizability density dyadic
 \begin{equation} \l{ealpha}
 \=\alpha = \le \=\eps_{\,a} - \=\eps_{\,b} \ri \. \les \, \=I + \=D \. \le \=\eps_{\,a} - \=\eps_{\,b} \ri \ris^{-1}.
 \end{equation}
 Two depolarization dyadics appear in Eqs.~\r{eMG} and \r{ealpha}. First,
 \begin{equation} \l{Ds_def}
 \=D_{\,s} = \=D^0_{\,s} + \=D^+_{\,s}
 \end{equation}
  which corresponds to a spherical Lorentzian cavity, with
  \begin{equation}
  \=D^0_{\,s} = \frac{1}{3 \eps_b} \=I
  \end{equation}
   being the depolarization contribution arising in the limit $\rho \to 0$ \c{ML_PiO}, and
   \begin{equation}
  \=D^+_{\,s}  = - \frac{\ko^2 \rho^2}{9 \epso} \le 3+ 2 i \ko \rho n_b \ri \=I
   \end{equation}
     being the depolarization contribution arising from nonzero $\rho$ \c{M_WRM}. Second,
   \begin{equation} \l{D_def}
 \=D = \=D^0 + \=D^+
 \end{equation}
  which corresponds to a pore whose shape and orientation are specified by $\=U$ per Eqs.~\r{rs_def} and \r{Udef},
 with
  \begin{equation}
  \=D^0 =  D^{0,\perp} \le \=I - \hat{\#z} \,\hat{\#z} \ri + D^{0,\parallel} \, \hat{\#z} \, \hat{\#z}
  \end{equation}
   being the depolarization contribution arising in the limit $\rho \to 0$ \c{ML_PiO}, and
   \begin{equation}
  \=D^+  = D^{+,\perp} \le \=I - \hat{\#z} \,\hat{\#z} \ri + D^{+,\parallel} \, \hat{\#z} \, \hat{\#z}
   \end{equation}
     being the depolarization contribution arising from nonzero $\rho$ \c{M_WRM}.
 Herein the depolarization dyadic components \c{M97,MW97}
 \begin{equation}
 \left. \l{D0_perp_parallel}
 \begin{array}{l}
 D^{0,\perp} = \displaystyle{\frac{1}{2 \tilde{\eps}_b^\perp \le U^\perp \ri^2} \les \frac{\tanh^{-1} \sqrt{1- \tilde{\gamma}}}{\sqrt{1- \tilde{\gamma}}} \le 1 + \frac{1}{\tilde{\gamma} - 1} \ri -
 \frac{1}{\tilde{\gamma} - 1} \, \ris } \vspace{6pt} \\
  D^{0,\parallel} = \displaystyle{ \frac{1}{\tilde{\eps}_b^\perp \le \tilde{\gamma} - 1 \ri \le U^\parallel \ri^2 } \le 1 -  \frac{\tanh^{-1} \sqrt{1- \tilde{\gamma}}}{\sqrt{1- \tilde{\gamma}}} \ri }
 \end{array}
 \right\}
 \end{equation}
   and \c{M_JNP}
 \begin{equation} \l{D+_perp_parallel}
 \left.
 \begin{array}{l}
 D^{+,\perp} = - \displaystyle{\frac{\ko^2 \rho^2}{8 \epso \le U^\perp \ri^2} \les \, \frac{\tilde{\gamma}}{1-\tilde{\gamma}} +
 i \frac{4 \ko \rho \sqrt{\tilde{\eps}^\perp}_b }{9 \sqrt{\epso}} \le 3 + \tilde{\gamma} \ri
 - \le \frac{\tilde{\gamma}}{1- \tilde{\gamma}} \ri^2 \sqrt{1- \tilde{\gamma}} \, \tanh^{-1} \sqrt{1- \tilde{\gamma}} \, \ris
 } \vspace{6pt} \\
 D^{+,\parallel} = - \displaystyle{\frac{\ko^2 \rho^2}{4 \epso \le U^\parallel \ri^2} \les \, i \frac{4 \ko \rho \sqrt{\tilde{\eps}_b^\perp}}{9 \sqrt{\epso}}  -
 \frac{1}{1 - \tilde{\gamma}} \le 1 - \frac{2 - \tilde{\gamma}}{1- \tilde{\gamma}} \sqrt{1- \tilde{\gamma}}  \, \tanh^{-1} \sqrt{1- \tilde{\gamma}} \ri
 \ris }
 \end{array}
 \right\},
 \end{equation}
   with $\tilde{\eps}^\perp_b = \eps^\perp_b / \le U^\perp \ri^2$ and  the dimensionless parameter $\tilde{\gamma} = \le U^\perp / U^\parallel \ri^2 \le  \eps^\parallel_b / \eps^\perp_b \ri$.
 Usually, the standard version of the Maxwell Garnett formalism is implemented wherein the approximation $\=D_{\,s} = \=D^0_{\,s}$
 and
 $\=D = \=D^0$
 is made; i.e., the nonzero size of the component pores is  neglected. Indeed, the standard version  of the Maxwell Garnett formalism is   implemented in the numerical studies presented in the following \S\ref{MG_sec}. However, in \S\ref{exMG_sec}, the nonzero size of the component pores
 is explicitly taken into account. This requires the extended version of the Maxwell Garnett formalism wherein the depolarization dyadics $\=D_{\,s}$ and $\=D$ are as given in Eqs.~\r{Ds_def} and \r{D_def}, respectively \c{M_JNP}.
Notice that the standard version is identical to the extended version implemented in the limit $\rho \to 0$.

\section{Numerical studies} \l{Num_sec}

We investigate the prospects for optical sensing in a fluid--infiltrated hyperbolic material by means of  numerical studies based on the homogenization approach described in \S\ref{MG_Sec}. The envisaged mechanism for sensing is delivered by the transition
 between propagating solutions and evanescent solutions
that emerges from corresponding dispersion relation for extraordinary plane waves. The transition
  occurs when the angle between the wave--vector direction and the optic axis of the infiltrated
 hyperbolic material is
   $ \le \pi/2 \ri - \theta_c$.
 Thus, in the following attention is focussed on the critical angle $\theta_c$ and its dependency on the refractive index $n_b$ of the fluid containing the analyte to be sensed.
 The derivative $d \theta_c / d n_b$ provides a measure of the sensitivity for
the envisaged method of optical sensing.

 Two qualitatively different scenarios are considered. In \S\ref{MG_sec}, the standard Maxwell Garnett formalism is used to investigate the effects of  porosity, pore shape and permittivity parameters of the porous hyperbolic material, as well as the refractive index of  the infiltrating fluid, on the sensitivity measure  $d \theta_c / d n_b$. In \S\ref{exMG_sec}, the extended Maxwell
  Garnett formalism is used to investigate the effect of  pore size on the sensitivity measure  $d \theta_c / d n_b$. The use of the extended Maxwell Garnett formalism results in a dissipative HCM, as may be appreciated from Eqs.~\r{D+_perp_parallel}. The dissipation introduced by such an extended homogenization formalism may be attributed to coherent scattering losses arising from the nonzero size of the pores. This phenomenon is a general feature of extended homogenization formalisms. Strictly, the corresponding HCM is not hyperbolic in the sense defined in \S\ref{Hyperbolic_sec} because its wave--vector components are complex--valued. However, provided that $\vert \, \mbox{Im} \lec \eps^{\perp,\parallel}_{hcm} \ric \vert \ll \vert \, \mbox{Re} \lec \eps^{\perp,\parallel}_{hcm} \ric \vert $, the real parts of the extraordinary  wave--vector components satisfy the dispersion relation
\begin{equation} \l{ex_exMG}
\frac{ \mbox{Re} \lec k_x \ric^2 +  \mbox{Re} \lec k_y \ric^2 }{ \mbox{Re} \lec \eps_{hcm}^\parallel \ric} +
\frac{\mbox{Re} \lec k_z \ric ^2}
{\mbox{Re} \lec \eps_{hcm}^\perp \ric}  =  \frac{\ko^2}{\epso} \,
\end{equation}
and the HCM may be regarded as being approximately hyperbolic.

\subsection{Standard Maxwell Garnett formalism} \l{MG_sec}

Suppose that the porous hyperbolic component material is specified by the
  permittivity parameters $\eps^\perp_a = 3.5 \epso$ and $\eps_a^\parallel = -2 \epso$ and the infiltrating fluid by the refractive index $n_b = 1.3$.
  The corresponding estimates of the HCM's permittivity parameters, as yielded by the standard Maxwell Garnett formalism,  are plotted against volume fraction $f_a$ in Fig.~\ref{fig2} with the relative shape parameter  $ (U^\parallel/U^\perp) \in \lec 0.75,  1.0,  1.25 \ric$.
  In conformity with Eq.~\r{eMG}, the volume fraction range is restricted to $0 < f_a < 0.3$.
  As $f_a$ increases, the HCM permittivity parameter $\eps^\perp_{hcm}$ increases approximately linearly from its starting point at $1.3^2 \epso$, and the rate of increase is largely independent of the value of $U^\parallel/U^\perp$. In contrast,  the HCM permittivity parameter $\eps^\parallel_{hcm}$ is highly sensitive to  the value of $U^\parallel/U^\perp$. As $f_a$ increases, $\eps^\parallel_{hcm}$
makes the transition from being positive--valued to being negative--valued~---~which corresponds the HCM making the transition from being non--hyperbolic to being hyperbolic~---~at $f_a = 0.07, 0.19$ and 0.27 for  $(U^\parallel/U^\perp) = 0.75, 1.0$ and 1.25, respectively.

The normalized wave--vector components of extraordinary plane waves are illustrated in Fig.~\ref{fig3} for the HCM whose permittivity parameters are displayed in Fig.~\ref{fig2}. Only the volume fractions $f_a \in \lec 0.1, 0.2, 0.3 \ric$ for $(U^\parallel/U^\perp) = 1.0$ are represented in Fig.~\ref{fig3}. At $f_a = 0.3$, the range of directions in which extraordinary plane waves can propagate is relatively large, with the critical angle being $\theta_c = 62.7^\circ$. At $f_a = 0.2$, this range of directions is much reduced with
the critical angle being $\theta_c = 78.8^\circ$. And at $f_a = 0.1$ the material is not  hyperbolic at all~---~the corresponding dispersion relation has a spheroidal representation in $(k_x, k_y, k_z )$ space.
\blue{Parenthetically,
the transition from spheroidal (or elliptical) to hyperbolic dispersion relations may be characterized by a sharp increase in  photon density of states \c{PNAS}, which can result in increased rates of spontaneous emission  from nearby emitters \c{Krishnamoorthy,Nature_nanotech}. Furthermore, for graphene-based hyperbolic materials, for example, this transition may be controlled by electrostatic biasing \c{JNP_graphene}.}

The derivative  $d \theta_c / d n_b$, which is adopted here as a measure of sensitivity for optical sensing, is plotted against volume fraction $f_a$ in Fig.~\ref{fig4}, for
the same component materials as were considered in Fig.~\ref{fig2},
with the relative shape parameter  $ (U^\parallel/U^\perp) \in \lec 0.75,  1.0,  1.25 \ric$.
 The presented ranges for $f_a$ are restricted to those ranges for which the HCM is hyperbolic; i.e., $f_a> 0.07$ for $(U^\parallel/U^\perp) = 0.75$,
\blue{$f_a> 0.19$} for $(U^\parallel/U^\perp) = 1.0$, and $f_a> 0.27$ for $(U^\parallel/U^\perp) = 1.25$.
For all values of $U^\parallel/U^\perp$,
the sensitivity measure $d \theta_c / d n_b$ increases rapidly as $f_a$ decreases towards its lower bound beyond which the HCM is no longer hyperbolic; and in this region of rapid increase, $d \theta_c / d n_b$ attains values in excess of 500 degrees per refractive index unit (RIU).
 Away from the region of rapid increase, for example at $f_a = 0.3$ for $ (U^\parallel/U^\perp) \in \lec 0.75,  1.0\ric$,
 $d \theta_c / d n_b$ attains values in the range 60--80  degrees per RIU.

The effect of the pore shape on the sensitivity for optical sensing is explored further in Fig~\ref{fig5}, wherein
the derivative  $d \theta_c / d n_b$ is plotted against the relative shape parameter  $U^\parallel/U^\perp$ for the volume fraction $f_a \in \lec 0.2, 0.25, 0.3 \ric$. As in Fig.~\ref{fig2}, the porous hyperbolic  material is specified by the
  permittivity parameters $\eps^\perp_a = 3.5 \epso$ and $\eps_a^\parallel = -2 \epso$ and the infiltrating fluid by the refractive index $n_b = 1.3$.
We see that the sensitivity measure $d \theta_c / d n_b$  increases uniformly,
 at  a fixed value of porosity,
 as the pores become more elongated in the direction of the optic axis  of the hyperbolic HCM. This general trend is independent of $f_a$, but generally higher values of $d \theta_c / d n_b$ are attained when $f_a$ is smaller.

Next we turn to the  refractive index $n_b$ of the analyte--containing fluid that infiltrates the porous hyperbolic material. Its bearing  upon the sensitivity measure $d \theta_c / d n_b$ is illustrated by Fig.~\ref{fig6}, wherein  $d \theta_c / d n_b$ is plotted against $n_b$ for  the volume fraction $f_a \in \lec 0.2, 0.25, 0.3 \ric$. The pores are assumed to be spherical for these plots; i.e., the relative shape parameter $ (U^\parallel/U^\perp) = 1.0$. The permittivity parameters of component material `a' are the same as for Fig.~\ref{fig2}.
In Fig.~\ref{fig6},
$d \theta_c / d n_b$  increases uniformly,
 at  a fixed value of porosity, as the refractive index $n_b$ increases. Furthermore, the values of $d \theta_c / d n_b$ become exceedingly large as $n_b$ increases towards its limiting value beyond which the HCM is no longer hyperbolic. As the volume fraction increases, this limiting value of $n_b$ occurs at higher values of $n_b$.

 The influence of the permittivity parameters of the porous hyperbolic
material upon the sensitivity measure $d \theta_c / d n_b$ is delineated in Fig.~\ref{fig7}. Therein, $d \theta_c / d n_b$  is plotted against $\eps^{\perp}_{a} / \epso$ and $\eps^{\parallel}_{a} / \epso$ for the volume fraction $f_a = 0.3$, relative shape parameter $ (U^\parallel/U^\perp) = 0.8$, and refractive index of the analyte--containing fluid $n_b = 1.3$.
The sensitivity measure decreases slightly as $\eps^\perp_a$ increases from $\epso$ to $ 4 \epso$, with the gradient of decent being greater at larger values of $\eps^\parallel_a$. On the other hand, the
sensitivity measure increases markedly as $\eps^\perp_a$ increases from $-2 \epso$ to $ -1.3 \epso$, with the gradient of ascent being greater at smaller values of $\eps^\perp_a$. Thus, the greatest values of $d \theta_c / d n_b$ are attained at the smallest values of $\eps^\perp_a$ and largest values of $\eps^\parallel_a$.

\subsection{Extended Maxwell Garnett formalism} \l{exMG_sec}

Now we consider the influence of pore size upon the sensitivity measure $d \theta_c / d n_b$. For this purpose, the extended Maxwell Garnett formalism is required. The real and imaginary parts of the HCM's permittivity parameters, as estimated by
the extended Maxwell Garnett formalism, are plotted against the normalized size parameter $\ko \rho$ in Fig.~\ref{fig8},
with the relative shape parameter \blue{ $ (U^\parallel/U^\perp) =   1.0$ and volume fraction $f_a \in \lec 0.2, 0.25,  0.3 \ric$}.
 As in Fig.~\ref{fig2}, the porous hyperbolic material is specified by the
  permittivity parameters $\eps^\perp_a = 3.5 \epso$ and $\eps_a^\parallel = -2 \epso$ and the infiltrating fluid by the refractive index $n_b = 1.3$. The real part of $\eps^\perp_{hcm}$ increases slightly as the pore size  increases, with the values of $\mbox{Re} \lec \eps^\perp_{hcm} \ric$ being only slightly influenced by the \blue{volume fraction}. In contrast, the real part of $\eps^\parallel_{hcm}$ increases markedly as the pore size  increases, with the values of $\mbox{Re} \lec \eps^\parallel_{hcm} \ric$ being considerably influenced by the \blue{volume fraction}. The imaginary parts of
$\eps^\perp_{hcm}$ and $\eps^\parallel_{hcm}$ both increase from zero as the size parameter increases, with the rate of increase being substantially greater for $\mbox{Im} \lec \eps^\parallel_{hcm} \ric $ than for $\mbox{Im} \lec \eps^\perp_{hcm} \ric$. These imaginary parts are only marginally effected by the  \blue{volume fraction}.

In order to implement the sensitivity measure $d \theta_c / d n_b$ for the $\rho > 0 $ regime, it is important to take into account the complex--valued nature of the HCM's permittivity parameters.
\blue{In regions where $\mbox{Re} \lec \eps^\parallel_{hcm} \ric < 0$
and $\mbox{Re} \lec \eps^\parallel_{hcm} \ric > 0$,
the magnitudes of the imaginary parts of $\eps^\perp_{hcm}$ and $\eps^\parallel_{hcm}$ are both generally very small in comparison to the magnitudes of the real parts of $\eps^\perp_{hcm}$ and $\eps^\parallel_{hcm}$, as  presented in Fig.~\ref{fig8}.
Exceptionally, when
 $\mbox{Re} \lec \eps^\parallel_{hcm} \ric $ is extremely close to zero (and is negative valued),  $\mbox{Im} \lec \eps^\parallel_{hcm} \ric$ can become relatively large compared to  $\mbox{Re} \lec \eps^\parallel_{hcm} \ric $
and consequently the characterization of the HCM's extraordinary dispersion relation per Eq.~\r{ex_exMG} is not appropriate. The regime where
$\mbox{Re} \lec \eps^\parallel_{hcm} \ric \approx 0$ is discussed further in Section~\ref{closing}.
Therefore, excluding the  regime where $\mbox{Re} \lec \eps^\parallel_{hcm} \ric \approx 0$,
 the HCM may be regarded as being approximately hyperbolic with the real parts of the extraordinary wave--vector components satisfying the
dispersion relation \r{ex_exMG}.} Accordingly, here we extend the definition \r{theta_c_def} of the critical angle, which marks the transition from propagating solutions to evanescent solutions emerging from the corresponding dispersion relation for extraordinary plane waves, to
\begin{equation} \l{tcd}
\theta_c = \tan^{-1} \sqrt{-  \frac{\mbox{Re} \lec \eps_{hcm}^\perp \ric }{ \mbox{Re} \lec \eps_{hcm}^\parallel \ric}}.
\end{equation}
That is, extraordinary plane waves  propagate in the approximately hyperbolic HCM
for directions given by $ 0 \leq \psi < \le \pi/2 \ri - \theta_c$, where $\psi = \cos^{-1} \le \mbox{Re} \lec k_z \ric / \mbox{Re} \lec k \ric \ri$ and the critical angle $\theta_c$ is as given in Eq.~\r{tcd}.

The derivative $d \theta_c / d n_b$ is plotted against the normalized size parameter $\ko \rho$ in Fig.~\ref{fig9}. Therein
the volume fraction $f_a \in \lec 0.2, 0.25,  0.3 \ric$,
 the relative shape parameter  $ (U^\parallel/U^\perp) = 1.0$ and the constitutive parameters of the component materials `a' and `b' are the same as for Fig.~\ref{fig8}. For all volume fractions considered, the sensitivity increases uniformly as the pore size increases. The rate of increase of $d \theta_c / d n_b$ is relatively modest at low values of $\rho$ but increases rapidly as $\rho$ increases. At a fixed value of $\rho$, the greatest values of $d \theta_c / d n_b$ are attained at the smallest values of $f_a$.

\section{Discussion} \l{closing}

A novel means of optical sensing has been  explored theoretically. It is based on a porous hyperbolic material which is infiltrated by a fluid containing an analyte to be sensed. Extraordinary plane waves can only propagate in the infiltrated hyperbolic material in directions enclosed by a cone which subtends the critical angle $\theta_c$ to the plane perpendicular to the optic axis of the hyperbolic material.
Numerical studies utilizing the Maxwell Garnett homogenization formalism have revealed that the critical angle $\theta_c$ is highly sensitive to the refractive index of the
 analyte--containing fluid, and this sensitivity can be strongly influenced by
  the permittivity parameters and porosity of the hyperbolic material, as well as
  the shape and size of its pores. Specifically, the sensitivity increases as:
   (i) the refractive index of the fluid increases; (ii) the negative--valued eigenvalue of the hyperbolic material's permittivity dyadic increases;
   (iii) the porosity increases; (iv) the pores become more elongated in the direction of optic axis;  and (v) the pores increase in size.

\blue{
The preceding analysis is based on the well--established Maxwell Garnett homogenization formalism  \c{MG_MOTL,ML_PiO}.
 The key assumptions underlying this approach are that the porosity of the hyperbolic host material is relatively high (i.e.,  $f_a \lessapprox 0.3$)
and that the pores are much smaller than the  wavelengths involved. Like most conventional approaches to homogenization, this approach is limited to materials which are spatially local. If the hyperbolic host material
were to possess a characteristic length scale which was  similar  to the
 wavelengths involved, then the effects of spatial nonlocality may become significant and an alternative homogenization approach may be needed \c{Chern_spatial_nonlocal}. However, if the hyperbolic host material were envisaged to be a homogenized composite material arising from  components made up of  particles much smaller than the wavelengths involved \c{Depine_MOTL}, for example, then the effects of spatial nonlocality are likely to be negligible.}

It is notable that exceeding high sensitivities, as gauged by the derivative $d \theta_c / d n_b$, can be achieved. For a HCM with
 permittivity parameter $\eps^\perp_{hcm} > 0$ (as investigated in \S\ref{Num_sec}),
the largest values of $d \theta_c / d n_b$ are attained when the negative--valued  HCM permittivity parameter $\eps^\parallel_{hcm}$ is very close to zero. Indeed, in the limit $\eps^\parallel_{hcm} \to 0^-$, the critical angle $\theta_c \to \pi/2$ and the derivative
$d \theta_c / d n_b$ becomes unbounded. \blue{However, when losses are taken into account, in the immediate vicinity of $\mbox{Re} \lec \eps^\parallel_{hcm} \ric \approx 0$, the magnitude of $\mbox{Re} \lec \eps^\parallel_{hcm} \ric$ can become relatively small compared
to that of  $\mbox{Im} \lec \eps^\parallel_{hcm} \ric$;  consequently the  characterization of the HCM's extraordinary dispersion relation per Eq.~(27) may be inappropriate. But
 even for values of $\eps^\parallel_{hcm}$ well away from the  immediate vicinity of $\mbox{Re} \lec \eps^\parallel_{hcm} \ric \approx 0$, the magnitudes of the sensitivity measure
 $d \theta_c / d n_b$ presented in Figs.~\ref{fig4}, \ref{fig5}, \ref{fig6}, \ref{fig7}
and \ref{fig9}
 compare favourably to sensitivities associated with optical sensing approaches based on surface--plasmon--polariton waves \c{Homola2003,AZL2,Scarano,ML_SJ_2012,SJJ_TGM_OC_2012}, Voigt waves \c{TGM_VW_1,TGM_VW_2}, as well as generic porous dielectric materials \c{TGM_AO}. We note that the issue of relatively large losses that arises in the parameter regime where $\mbox{Re} \lec \eps^\parallel_{hcm} \ric \approx 0$ could, in principle, be addressed by the use of active component materials \c{Gain_ENZ}.}

 The numerical results reported here bode well for the prospects of harnessing porous  hyperbolic materials as platforms for optical sensing, and thereby adds to the already long list of potential  applications for these exotic materials \c{Kivshar}.

\vspace{15mm}
\blue{
\noindent{\bf Acknowledgement.  } The author acknowledges the support of EPSRC grant EP/M018075/1.}

\begin{figure}[!ht]
\centering
\includegraphics[width=4.5in]{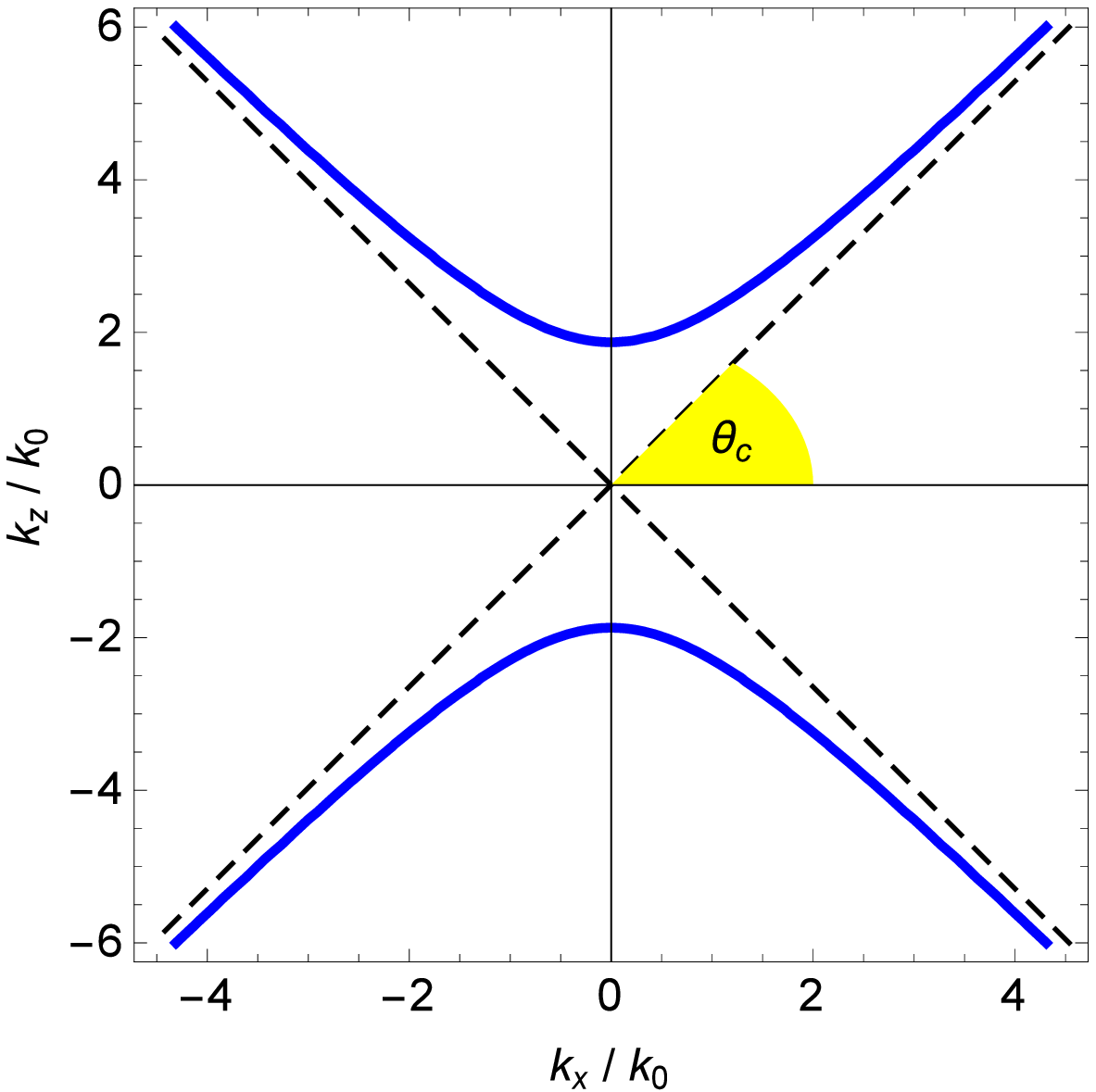}
 \caption{Normalized wave--vector components $k_x / \ko$ and $k_z / \ko$ of extraordinary plane waves for the permittivity parameters $\eps^\perp = 3.5 \epso$ and $\eps^\parallel = -2 \epso$.
 The $k_y = 0 $ plane is represented, for which
 the wave--vector components  are bounded by the cone $k_z^2 =  k_x^2 \, \tan^2  \theta_c  $, where $\theta_c = \tan^{-1} \sqrt{- \eps^\perp /\eps^\parallel }$.
  \l{fig1}}
\end{figure}

\newpage

\begin{figure}[!ht]
\centering
\includegraphics[width=4.5in]{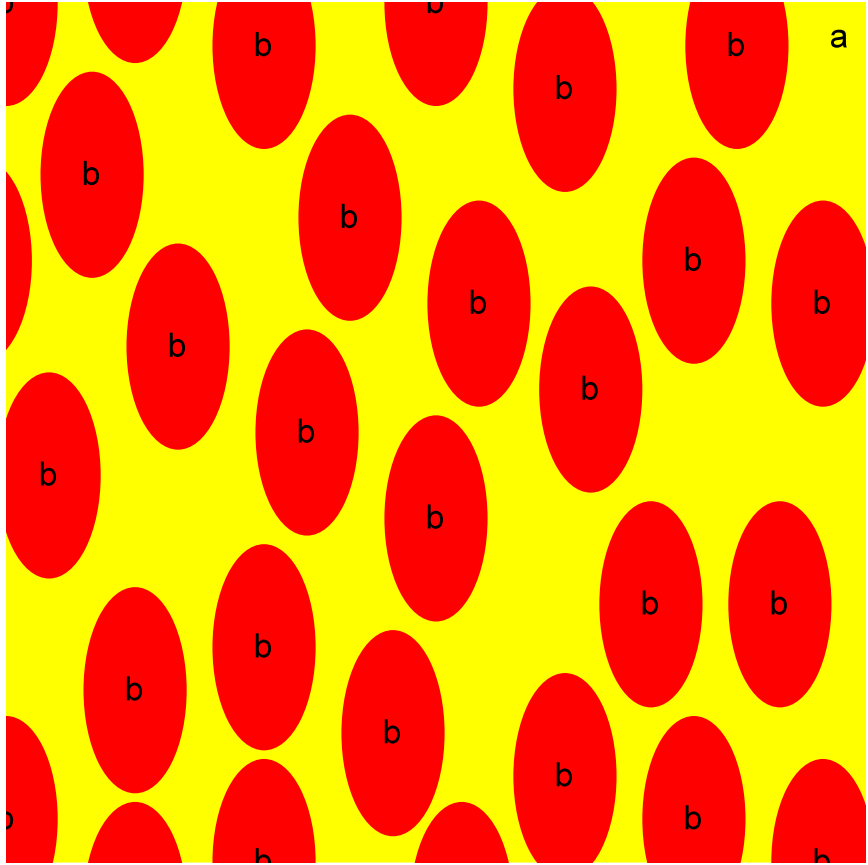}
 \caption{\blue{Schematic diagram showing the composite material to be homogenized using the Maxwell Garnett formalism.  A highly-porous hyperbolic host material (labelled `a') contains spheroidal pores which are filled with
   a fluid (labelled `b'). The electromagnetically-small pores all have the same orientation but they  are randomly
distributed.}
  \l{figx}}
\end{figure}

\newpage

\begin{figure}[!ht]
\centering
\includegraphics[width=4.5in]{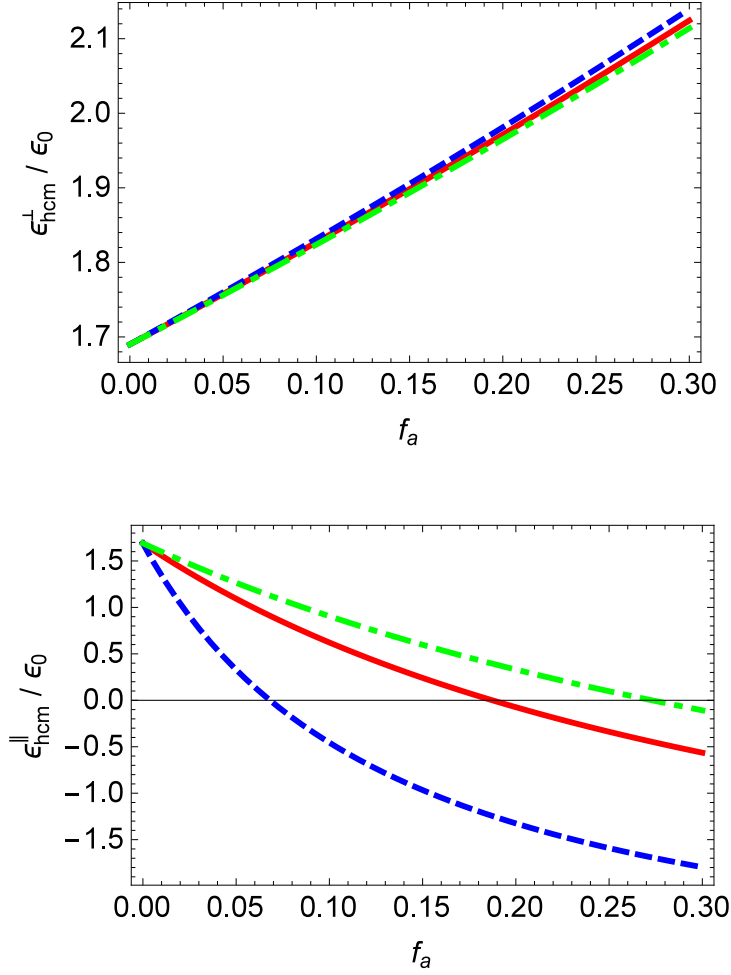}
 \caption{Maxwell Garnett estimates of $\eps^\perp_{hcm} / \epso$  and $\eps^\parallel_{hcm} / \epso$   plotted versus  volume fraction $f_a \in \le 0, 0.3 \ri$
 for the relative shape parameter $ (U^\parallel/U^\perp) = 0.75$ (blue, dashed curves), 1.0 (red, solid curves) and 1.25 (green, broken dashed curves),
 refractive index $n_b = 1.3$, and permittivity parameters $\eps^\perp_a = 3.5 \epso$ and $\eps_a^\parallel = -2 \epso$.
  \l{fig2}}
\end{figure}

\newpage

\begin{figure}[!ht]
\centering
\includegraphics[width=6.5in]{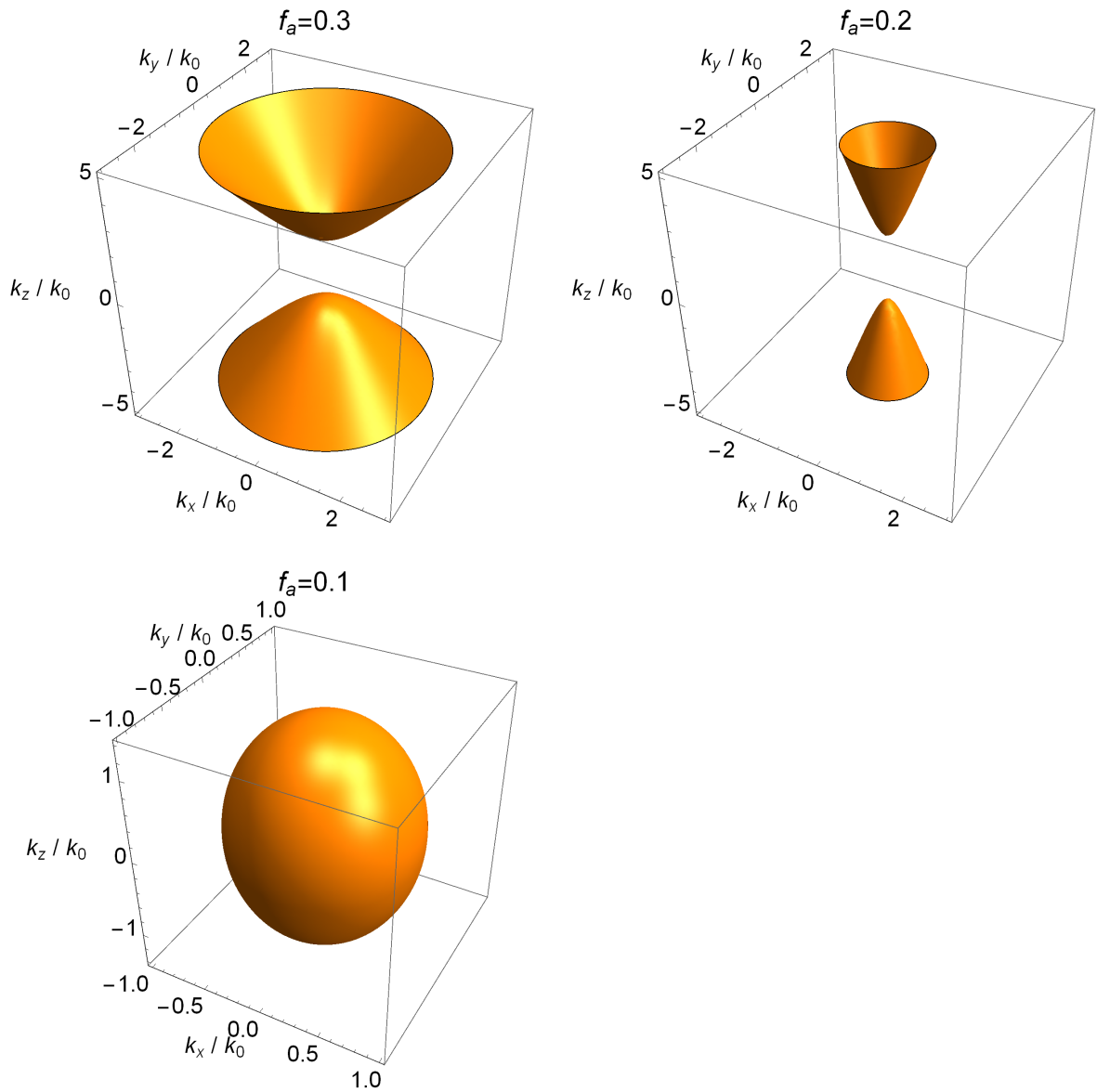} 
 \caption{Normalized wave--vector components of extraordinary plane waves for  volume fractions $f_a \in \lec 0.1, 0.2, 0.3 \ric$, with the
  relative shape parameter $ (U^\parallel/U^\perp) = 1$,
 refractive index $n_b = 1.3$, and permittivity parameters $\eps^\perp_a = 3.5 \epso$ and $\eps_a^\parallel = -2 \epso$.
  \l{fig3}}
\end{figure}

\newpage

\begin{figure}[!ht]
\centering
\includegraphics[width=4.5in]{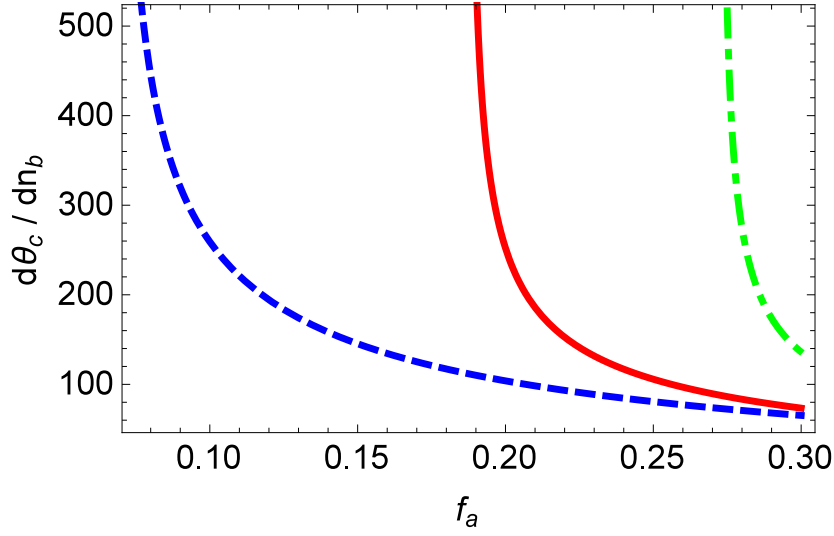}
 \caption{The derivative  $d \theta_c / d n_b$ (in degrees per RIU) plotted versus volume fraction $f_a $
 for the relative shape parameter $(U^\parallel/U^\perp) = 0.75$ (blue, dashed curve), 1.0 (red, solid curve) and 1.25 (green, broken dashed curve),
 refractive index $n_b = 1.3$, and permittivity parameters $\eps^\perp_a = 3.5 \epso$ and $\eps_a^\parallel = -2 \epso$.
  \l{fig4}}
\end{figure}

\newpage

\begin{figure}[!ht]
\centering
\includegraphics[width=4.5in]{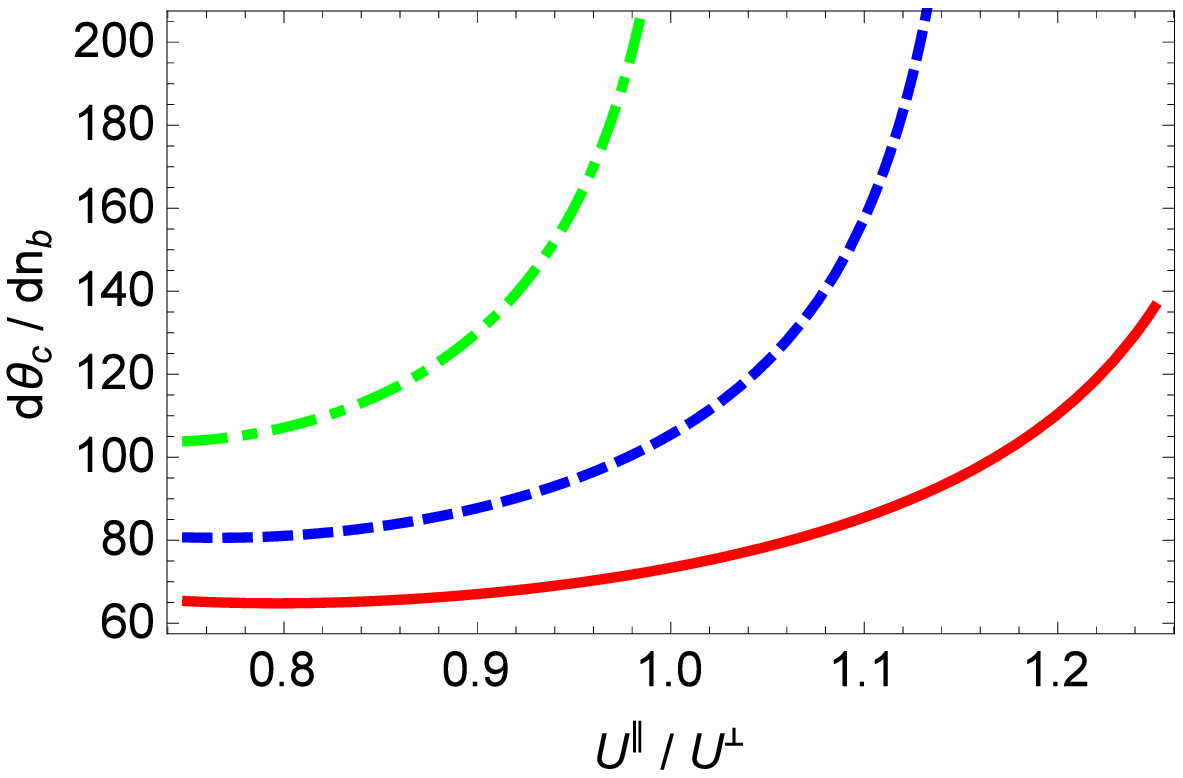}
 \caption{The derivative  $d \theta_c / d n_b$ (in degrees per RIU)
  plotted versus relative shape parameter $U^\parallel/U^\perp$ for the volume fraction $f_a = 0.3$ (red, solid curve),
   0.25 (blue, dashed curve),
   and 0.2 (green, broken dashed curve),
 refractive index $n_b = 1.3$, and permittivity parameters $\eps^\perp_a = 3.5 \epso$ and $\eps_a^\parallel = -2 \epso$.
  \l{fig5}}
\end{figure}

\newpage

\begin{figure}[!ht]
\centering
\includegraphics[width=4.5in]{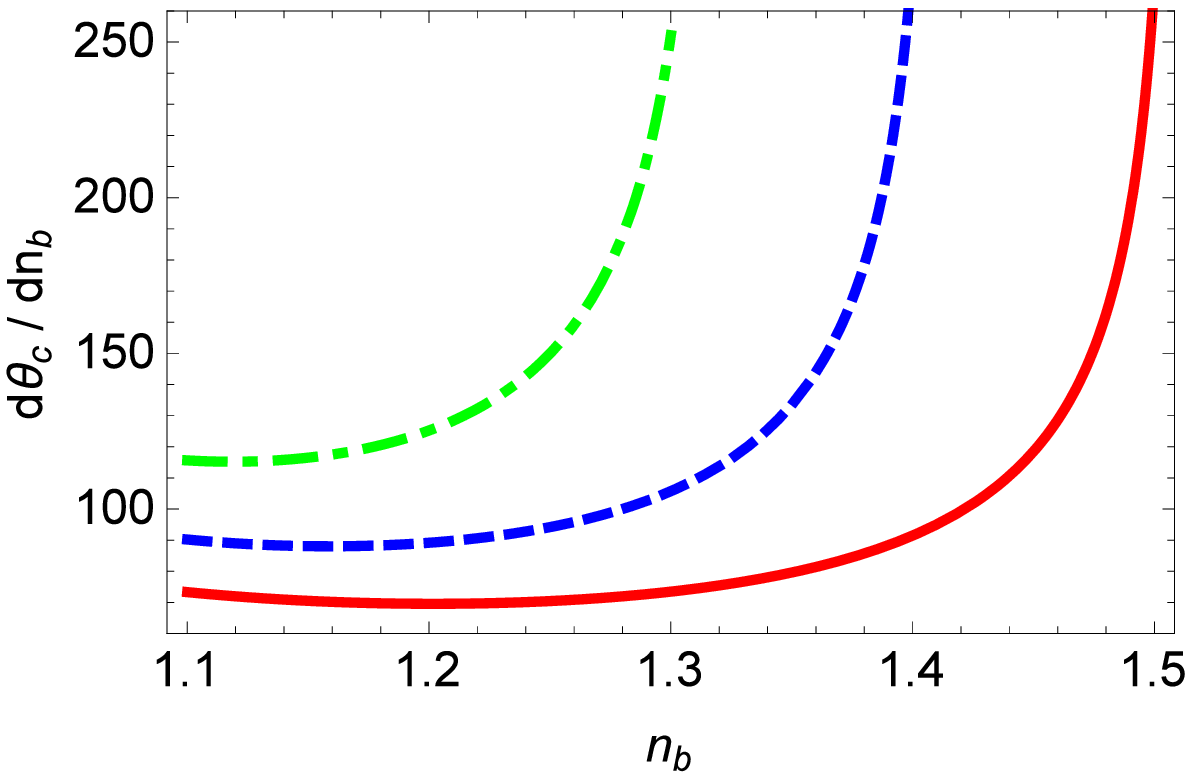}
 \caption{The derivative  $d \theta_c / d n_b$ (in degrees per RIU)
  plotted versus refractive index $n_b $ for the volume fraction $f_a = 0.3$ (red, solid curve),
   0.25 (blue, dashed curve),
   and 0.2 (green, broken dashed curve), relative shape parameter $\le U^\parallel/U^\perp \ri = 1$,
  and permittivity parameters $\eps^\perp_a = 3.5 \epso$ and $\eps_a^\parallel = -2 \epso$.
  \l{fig6}}
\end{figure}

\newpage

\begin{figure}[!ht]
\centering
\includegraphics[width=4.5in]{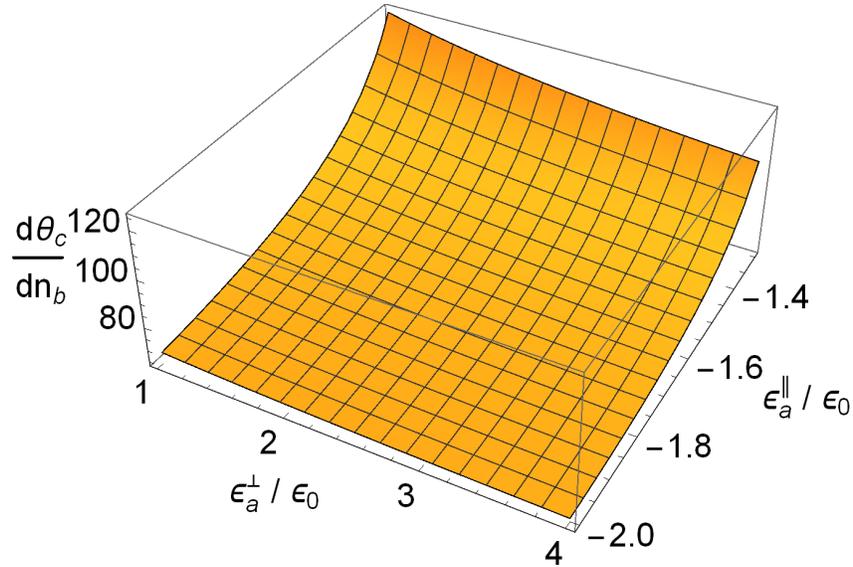}
 \caption{The derivative  $d \theta_c / d n_b$ (in degrees per RIU)
  plotted versus
 relative permittivity parameters $\eps^\perp_a / \epso$ and $\eps_a^\parallel / \epso$,  for the volume fraction $f_a = 0.3$,
   refractive index $n_b = 1.3$, and
   relative shape parameter $\le U^\parallel/U^\perp \ri = 0.8$.
  \l{fig7}}
\end{figure}

\newpage

\begin{figure}[!ht]
\centering
\includegraphics[width=6.5in]{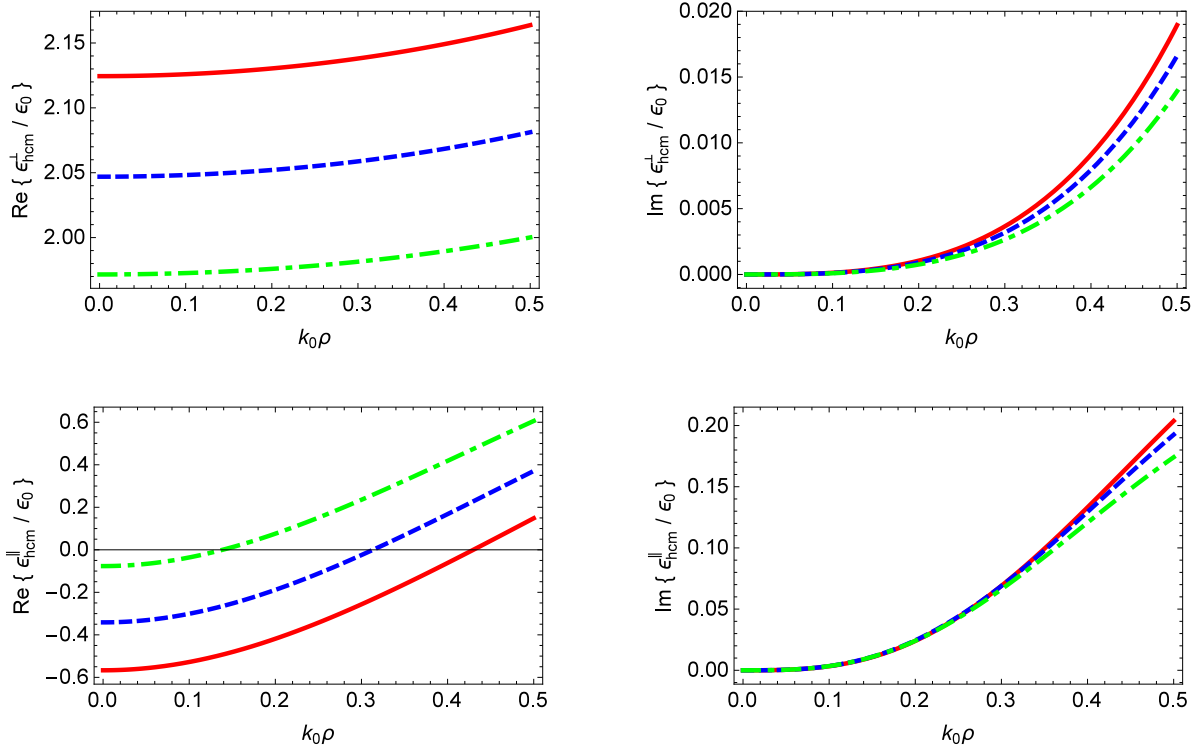} 
 \caption{Extended Maxwell Garnett estimates of the real and imaginary parts of $\eps^\perp_{hcm} / \epso$  and $\eps^\parallel_{hcm} / \epso$   plotted versus   relative size parameter $\ko \rho$
 for
  the volume fraction $f_a = 0.3$ (red, solid curve),
   0.25 (blue, dashed curve),
   and 0.2 (green, broken dashed curve),  relative shape parameter  $ (U^\parallel/U^\perp) = 1.0$,
 refractive index $n_b = 1.3$, and permittivity parameters $\eps^\perp_a = 3.5 \epso$ and $\eps_a^\parallel = -2 \epso$.
  \l{fig8}}
\end{figure}

\newpage

\begin{figure}[!ht]
\centering
\includegraphics[width=4.5in]{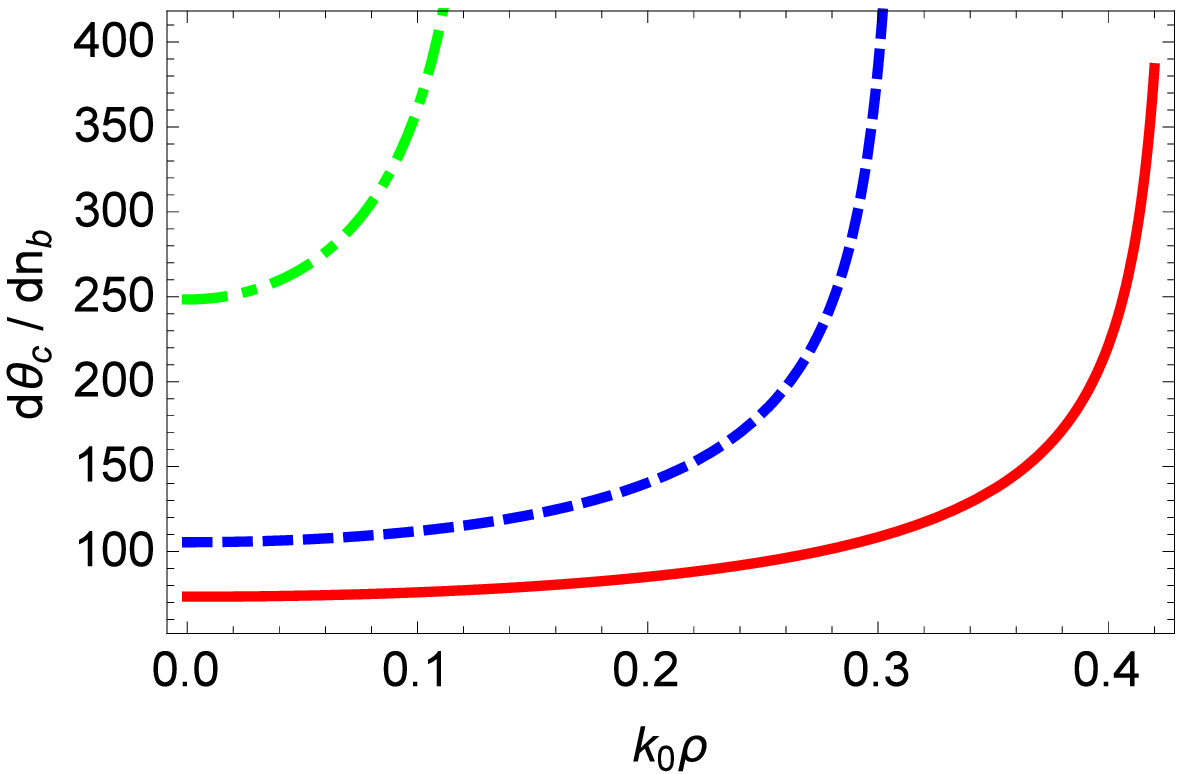}
 \caption{The derivative  $d \theta_c / d n_b$ (in degrees per RIU)
  plotted versus relative size parameter $\ko \rho$ for the volume fraction $f_a = 0.3$ (red, solid curve),
   0.25 (blue, dashed curve),
   and 0.2 (green, broken dashed curve),  relative shape parameter  $ (U^\parallel/U^\perp) = 1.0$,
 refractive index $n_b = 1.3$, and permittivity parameters $\eps^\perp_a = 3.5 \epso$ and $\eps_a^\parallel = -2 \epso$.
  \l{fig9}}
\end{figure}


\vspace{15mm}

\begin{thebibliography}{99}

\bibitem{Kivshar}
A. Poddubny, I. Iorsh, P. Belov, and Y. Kivshar, ``Hyperbolic metamaterials,"
\emph{Nature Photon.} {\bf 7}, 958--967  (2013).

\bibitem{Smith04}
D.R. Smith, P. Kolinko,  and D. Schurig, ``Negative refraction in indefinite media," \emph{J. Opt. Soc. Am. B} {\bf 21}, 1032--1043 (2004). 

\bibitem{Schilling}
J. Schilling,
``Uniaxial metallo-dielectric metamaterials with scalar positive permeability,"
\emph{Phys. Rev. E} {\bf 74}, 046618 (2006).

\blue{
\bibitem{Liu_Science_2007}
Z. Liu, H. Lee, Y. Xiong, C. Sun, and X. Zhang,
``Far-field optical hyperlens magnifying
sub-diffraction-limited objects,"
\emph{Science} {\bf 315}, 1686 (2007).}


\bibitem{Li_JAP}
G.X. Li, H.L. Tam, F.Y. Wang, and K.W. Cheah, ``Superlens from complementary anisotropic metamaterials," \emph{J. Appl. Phys.} {\bf
102}, 116101 (2007). 

\bibitem{Lemoult}
F. Lemoult, G. Lerosey, J. de Rosny, and M. Fink,
``Resonant metalenses for breaking the diffraction barrier,"
\emph{Phys. Rev. Lett.} {\bf 104}, 203901  (2010).


\bibitem{Hyperbolic_Planck}
Y. Guo, C.L.  Cortes, S. Molesky, and Z.  Jacob,  ``Broadband super--Planckian thermal emission from
hyperbolic metamaterials," \emph{ Appl.  Phys. Lett.} {\bf 101},
131106 (2012). 

\bibitem{Hyperbolic_Guo_OE}
Y. Guo and  Z. Jacob, ``Thermal hyperbolic metamaterials," \emph{Opt. Express} {\bf 21},
15014–-15019 (2013). 

\bibitem{Smoly}
I.I. Smolyaninov and E.E. Narimanov,
``Metric signature transitions in optical metamaterials,"
\emph{Phys. Rev. Lett.} {\bf 105}, 067402 (2010). 

\bibitem{ML_PRB}  T.G.   Mackay   and   A. Lakhtakia,
 ``Towards a  realization of Schwarzschild--(anti--)de Sitter spacetime as a particulate metamaterial,"
   \emph{Phys. Rev. B} {\bf 83}, 195424 (2011). 

\bibitem{Depine_NJP}
R.A. Depine and A. Lakhtakia,  ``Diffraction by a grating made of a uniaxial dielectric--magnetic medium exhibiting negative
refraction," \emph{New J. Phys.} {\bf 7}, 158 (2005).

\bibitem{Shen_PRB}
L. Shen, T.--J. Yang, and Y.--F. Chau, ``Effect of internal period on the optical dispersion of indefinite-medium materials,"
 \emph{Phys. Rev. B} {\bf 77}, 205124 (2008).

\bibitem{Orlov}
A.V. Chebykin, A.A. Orlov, C.R. Simovski, Yu.S. Kivshar, and P.A. Belov,
``Nonlocal effective parameters of multilayered metal--dielectric metamaterials,"
\emph{Phys. Rev. B} {\bf 86}, 115420 (2012).

\bibitem{Wire}
C.R. Simovski, P.A.  Belov, A.V.  Atrashchenko, Y.S. Kivshar,
``Wire metamaterials: physics and applications," \emph{Adv. Mater.} {\bf 24}, 4229–-4248 (2012).

\bibitem{Depine_MOTL}
T.G. Mackay,  A. Lakhtakia, and R.A. Depine,
``Uniaxial dielectric media with hyperbolic dispersion relations,"
\emph{ Microw.
  Opt. Technol. Lett.} {\bf 48}, 363--367 (2006).

\bibitem{Graphite}
J. Sun, J. Zhou, B. Li, and F. Kang,
``Indefinite permittivity and negative refraction in natural material: graphite,"
\emph{Appl. Phys. Lett.} {\bf 98}, 101901 (2011).


\bibitem{Smith03}
D.R. Smith  and D. Schurig, ``Electromagnetic wave propagation in media with indefinite permittivity and permeability tensors,"
\emph{Phys. Rev. Lett.}  {\bf 90}, 077405 (2003). 


\bibitem{Noginov}
M.A. Noginov, ``Compensation of surface plasmon loss by gain in dielectric medium," \emph{J. Nanophoton.} {\bf 2},  021855 (2008).

\bibitem{Vincenti}
M.A. Vincenti, D. de Ceglia, V. Rondinone, A. Ladisa, A. D'Orazio, M.J. Bloemer, and M. Scalora,
``Loss compensation in metal--dielectric structures in
negative--refraction and super--resolving regimes," \emph{Phys. Rev. A} {\bf 80}, 053807 (2009).


\bibitem{M_JNP}
T.G.  Mackay,
 ``On extended homogenization formalisms for
nanocomposites,"
  \emph{J.  Nanophoton.} {\bf 2},  021850 (2008).

\bibitem{MG_MOTL} W.S. Weiglhofer, A.  Lakhtakia, and B.   Michel,
  ``Maxwell Garnett and Bruggeman formalisms for a particulate composite with bianisotropic host medium," \emph{Microw.
Opt. Technol. Lett.} {\bf 15}, 263--266 (1997).
Corrections:  {\bf 22} , 221 (1999). 


\bibitem{ML_PiO}
T.G. Mackay and A. Lakhtakia, ``Electromagnetic fields in linear bianisotropic
mediums," \emph{Prog. Optics} {\bf 51}, 121--209 (2008).

\bibitem{BW}
M. Born and E. Wolf, \emph{Principles of Optics, 7th (expanded) edn},
Cambridge University Press, Cambridge, UK, 1999.

\bibitem{Chen}
H.C. Chen, \emph{Theory of Electromagnetic Waves}, McGraw--Hill,
New York, NY, USA, 1983.

\bibitem{M_WRM}
T.G.   Mackay,
 ``Depolarization volume and correlation length in the homogenization
of anisotropic dielectric composites," \emph{Waves  Random Media} {\bf
14}, 485--498 (2004). Corrections: \emph{Waves Random Complex Media} {\bf
16}, 85 (2006).



\bibitem{M97}
B. Michel,   ``A Fourier space approach to the pointwise singularity
of an anisotropic dielectric medium," \emph{Int. J. Appl. Electromagn.
Mech.} {\bf 8}, 219--227 (1997).

\bibitem{MW97}
B. Michel  and W.S. Weiglhofer,  ``Pointwise singularity of dyadic
Green function in a general bianisotropic medium," \emph{Arch.
Elekron. \"Ubertrag.} {\bf 51}, 219--223 (1997). Corrections: {\bf 52}, 31 (1998).

\blue{
\bibitem{PNAS}
G.V. Naik, B. Saha, J. Liu, S.M. Saber, E.A. Stach, J.M.K. Irudayaraj, T.D. Sands,
V.M. Shalaev, and A. Boltasseva,
``Epitaxial superlattices with titanium nitride as
a plasmonic component for optical
hyperbolic metamaterials," \emph{PNAS} {\bf 111}
7546–7551  (2014).
}

\blue{
\bibitem{Krishnamoorthy}
H.N.S. Krishnamoorthy, Z. Jacob, E. Narimanov,
I. Kretzschmar, and V.M. Menon,
``Topological transitions in metamaterials,"
\emph{Science} {\bf 336}, 205--209 (2012).}

\blue{
\bibitem{Nature_nanotech}
D. Lu, J.J. Kan, E.E. Fullerton, and Z. Liu,
``Enhancing spontaneous emission rates of
molecules using nanopatterned multilayer
hyperbolic metamaterials,"
\emph{Nature Nanotech.} {\bf 9} 48--53 (2014).
}

\blue{
\bibitem{JNP_graphene}
M.A.K. Othman, C, Guclu, and F. Capolino,
``Graphene–-dielectric composite metamaterials: evolution from elliptic to hyperbolic wavevector dispersion and the transverse epsilon-near-zero condition,"
\emph{J. Nanophoton.} {\bf 7} , 073089 (2013). }

\blue{
\bibitem{Chern_spatial_nonlocal}
R.--L. Chern, ``Spatial dispersion and nonlocal effective
permittivity for periodic layered
metamaterials," \emph{Opt. Express} {\bf 21}, 16514--16527 (2013).}


\bibitem{Homola2003}
J. Homola, ``Present and future of surface plasmon resonance
biosensors," \emph{Anal. Bioanal. Chem.} {\bf 377},  528--539
(2003).

\bibitem{AZL2}
I. Abdulhalim, M. Zourob, and A. Lakhtakia, ``Surface plasmon
resonance for biosensing: A mini-review," {\it Electromagnetics}
{\bf 28}, 214--242 (2008).

\bibitem{Scarano}
S. Scarano, M. Mascini, A.P.F. Turner, and M. Minunni, ``Surface
plasmon resonance imaging for affinity--based biosensors,"
\emph{Biosens. Bioelectron.} {\bf 25}, 957--966 (2010).


\bibitem{ML_SJ_2012}  T.G. Mackay and  A. Lakhtakia,
 ``Modeling  chiral sculptured thin films as platforms for
surface--plasmonic--polaritonic optical sensing,"  \emph{IEEE
Sensors J.} {\bf 12}, 273--280 (2012).

\bibitem{SJJ_TGM_OC_2012} S.S. Jamaian and  T.G. Mackay,
 ``On columnar thin films as platforms for
surface-plasmonic-polaritonic optical sensing: higher-order
considerations," \emph{Opt. Commun.} {\bf 285}, 5535--5542 (2012).

\bibitem{TGM_VW_1}
   T.G. Mackay,
 ``On the sensitivity of directions which support Voigt wave propagation in infiltrated biaxial dielectric materials,"
    \emph{J. Nanophoton.} {\bf 8}, 083993 (2014).

\bibitem{TGM_VW_2}
T.G. Mackay,
 ``Controlling Voigt waves by the Pockels effect,"
   \emph{J. Nanophoton.}  {\bf 9},  093599 (2015).

\bibitem{TGM_AO}
T.G. Mackay, ``On the sensitivity of generic porous optical sensors," \emph{Appl. Optics} {\bf 51}, 2752--2758 (2012).

\blue{
\bibitem{Gain_ENZ}
S. Campione, D. de Ceglia, M.A. Vincenti, M. Scalora, and F. Capolino,
``Electric field enhancement in $\eps$-near-zero slabs under TM-polarized oblique incidence,"
\emph{Phys. Rev. B} {\bf 87}, 035120 (2013).
}

\end{thebibliography}
\end{document}